\documentclass[nofootinbib,notitlepage,superscriptaddress,10pt,aps,pra,twocolumn]{revtex4-2}

\usepackage{xcolor}
\usepackage[utf8x]{inputenc}
\usepackage{amsmath}
\usepackage{amssymb}
\usepackage{graphicx}
\usepackage[normalem]{ulem}
\usepackage{epsf,epsfig}
\usepackage{psfrag}
\usepackage{color}
\usepackage{multirow}
\usepackage{diagbox}
\usepackage{url}

\makeatletter
\newcommand\stroke[1]{\mathpalette\stroke@aux{#1}}
\def\stroke@aux#1#2{%
  \ooalign{%
    \hfil$#1^{\;\, \_\hspace{-0.05cm}\_}$\hfil\cr
    \hfil$#1#2$\hfil\cr
  }%
}
\makeatother

\begin{document}

\title{Could quantum gravity slow down neutrinos?}

\author{Giovanni Amelino-Camelia}
\affiliation{Dipartimento di Fisica Ettore Pancini, Universit\`a di Napoli “Federico II”, Complesso Univ. Monte S. Angelo, I-80126
Napoli, Italy}
\affiliation{INFN, Sezione di Napoli}
\author{Maria Grazia Di Luca}
\affiliation{Scuola Superiore Meridionale, Largo S. Marcellino 10, I-80138 Napoli, Italy}
\affiliation{Dipartimento di Fisica Ettore Pancini, Universit\`a di Napoli “Federico II”, Complesso Univ. Monte S. Angelo, I-80126
Napoli, Italy}
\affiliation{INFN, Sezione di Napoli}
\author{Giulia Gubitosi}
\affiliation{Dipartimento di Fisica Ettore Pancini, Universit\`a di Napoli “Federico II”, Complesso Univ. Monte S. Angelo, I-80126
Napoli, Italy}
\affiliation{INFN, Sezione di Napoli}
\author{Giacomo Rosati}
\affiliation{Institute of Theoretical Physics, University of Wroc\l{}aw, Pl. Maksa Borna 9, Pl-50-204 Wroc\l{}aw, Poland}
\author{Giacomo D'Amico}
\affiliation{Department for Physics and Technology, University of Bergen,  NO-5020 Bergen, Norway}

\begin{abstract}
In addition to its implications for astrophysics, the hunt for GRB neutrinos could also be significant in quantum-gravity research, since they are excellent probes of the microscopic fabric of spacetime.
Some previous studies based on IceCube neutrinos had found intriguing  preliminary evidence that some of them might be GRB neutrinos with travel times affected by quantum properties of spacetime, with the noticeable feature that quantum spacetime would slow down some of the neutrinos while others would be sped up. Recently the IceCube collaboration revised significantly the estimates of the direction of observation of their neutrinos, and we here investigate how the corrected directional information affects the results of the previous quantum-spacetime-inspired analyses. We find that there is now no evidence for neutrinos sped up by quantum-spacetime properties, whereas the evidence for neutrinos slowed down by quantum spacetime is even stronger than previously found. Our most conservative estimates find a false alarm probability of less than 1\% for these "slow neutrinos", providing motivation for future studies on larger data samples.
\end{abstract}

\maketitle
Over the last
few years one of the most studied candidate effects  of quantum gravity has been in-vacuo dispersion,
an energy dependence of the speed
of ultrarelativistic  particles
(see {\it e.g.}
Refs.~\cite{gacLRR,jacobpiran,gacsmolin,grbgac,gampul,urrutia,gacmaj,myePRL,gacGuettaPiran,steckerliberati} and references therein).
This effect could also lead to observably large manifestations, even if,
as it appears to be safe to
assume \cite{gacLRR,jacobpiran,gacsmolin,grbgac,gampul,urrutia,gacmaj}, its characteristic length scale turns out to be of the order of the
minute Planck length,  or anyway not much larger than that.
Observations
of GRBs \cite{gacLRR,jacobpiran,gacsmolin,grbgac}, which emit (nearly-)simultaneously
photons of different energies and (probably \cite{waxbig,meszabig,dafnebig,otherbig}) neutrinos,
could be well suited for finding a manifestation of the novel energy dependence of the speed.

Some of us were involved in some studies~\cite{gacGuettaPiran,Ryan,RyanLensing,NatureAstro,MaNeutriniFirst,HuangLiMaNeutriniNearTeV}
of IceCube neutrino data which produced intriguing results:
these studies compared observation times and directions of GRBs with
those of IceCube neutrinos, finding preliminary statistical evidence of the fact that some
of those neutrinos would be GRB neutrinos receiving a contribution to their travel times from
quantum-gravity-induced in-vacuo dispersion.
A potential weak point of those analyses was that they found
comparable statistical significance for neutrinos being slowed down
and neutrinos being sped up by in-vacuo dispersion:
the most popular quantum-gravity intuition is that~\cite{GAClivingReview} all particles with half-integer spin should be affected by the same effect,
though a specific model with breakdown of relativistic invariance can accommodate~\cite{SteckerLiberatiNeutrini} for an effect which has opposite sign for the two helicities
of the neutrino (a scenario which might here be relevant since we have no helicity information for IceCube neutrinos).

Recently the IceCube collaboration revised significantly~\cite{IceCubeNewDirections} their estimates of the direction of observation of their neutrinos, and we here investigate how the corrected directional information affects the results of the previous quantum-gravity-inspired analyses.
We start by quickly reviewing the formalization and parametrization of the model already used
in the previous studies of Refs.~\cite{Ryan,RyanLensing,NatureAstro,MaNeutriniFirst,HuangLiMaNeutriniNearTeV}:
\begin{equation}
\Delta t = \eta D(1) \frac{\mathcal{K} (E,z)}{M_{P}}   \,
\label{mainnewone}
\end{equation}
where $\Delta t$ is the
contribution to the travel time of the neutrino from
quantum-gravity-induced in-vacuo dispersion,
$\eta$ is a dimensionless parameter to be determined experimentally, $M_{P}$ denotes
the Planck scale ($\sim 10^{28}eV$, inverse of the Planck length),
$D(z)$ is a function of the redshift $z$ of the GRB
associated to the neutrino\footnote{$\Omega_\Lambda$, $H_0$ and $\Omega_m$ denote, as usual,
respectively the cosmological constant, the Hubble parameter and the matter fraction, for which we take the values given in Ref.~\cite{Planck:2018vyg}; $D(1)$ in Eq.(\ref{mainnewone}) is of course $D$ evaluated for redshift of 1.}
$$D(z) = \int_0^z d\zeta \frac{(1+\zeta)}{H_0\sqrt{\Omega_\Lambda + (1+\zeta)^3 \Omega_m}} \, ,$$
and $\mathcal{K}$ is a function of $D(z)$ and of the energy $E$ of the
neutrino: $\mathcal{K} (E,z) \equiv E D(z)/D(1) \, .$

In the data analysis $\Delta t$ will be estimated
in terms of the difference in times of observation\footnote{If there were no quantum-gravity effects, a GRB neutrino would
be observed (nearly-)simultaneously with the GRB that produced it. We attribute the whole of the time-of-arrival difference
to the quantum-gravity-induced $\Delta t$ of the neutrino, since the photons composing the GRB signal are of much lower energies than our neutrinos
and the effect we are studying depends linearly on energy (so any $\Delta t$ of the photons would be negligible with respect to the $\Delta t$ of the neutrino).}
 between a GRB and a neutrino
which can be tentatively described as produced in coincidence with that GRB.
Our first task is to find these "GRB-neutrino candidates"~\cite{Ryan,RyanLensing,NatureAstro,MaNeutriniFirst,HuangLiMaNeutriniNearTeV}:
neutrinos whose direction is compatible with a GRB direction and whose energy and time
of observation render them compatible, according to Eq.(\ref{mainnewone}),
with the time of observation of the GRB.
As it was clarified in detail
in Refs.~\cite{Ryan,RyanLensing,NatureAstro,MaNeutriniFirst,HuangLiMaNeutriniNearTeV}
(and will be clear also from the analysis reported here below), for none of our GRB-neutrino candidates we can be certain that it is indeed a GRB neutrino: the correspondence between direction and time of observation can of course also occur accidentally and for the single association between GRB and neutrino our strategy of analysis will inevitably be inconclusive.  However, if such associations are numerous enough (and "significant enough", see below) we could end up in the position of being certain that our data sample contains some GRB neutrinos whose propagation was affected by quantum gravity even though we would not be able to establish which of our GRB-neutrino candidates actually are GRB neutrinos.

Since one of our main goals is to weigh the impact on this sort of quantum-gravity-motivated analysis of the
revised estimates~\cite{IceCubeNewDirections} of the direction of observation of IceCube neutrinos, for this study we use the same data sample and the same criteria for the selection of GRB-neutrino candidates already used in the previous studies of Refs.~\cite{Ryan,RyanLensing,NatureAstro,MaNeutriniFirst,HuangLiMaNeutriniNearTeV}, where the interested reader will find them described in detail.
We here just stress that the analyses are confined to  neutrinos with energy between 60 and 500 TeV and one has a GRB-neutrino candidate if the GRB and the neutrino were observed at times that differ by no more than 3 days and
the angular distance between the direction of the neutrino and the direction of the GRB is within a $3\sigma$ region,
with $\sigma=\sqrt{\sigma^2_{\text{GRB}}+\sigma^2_{\nu}}$
(with $\sigma_{\text{GRB}}$
 and $\sigma_\nu$ denoting, respectively, the directional uncertianties for the GRB and for the neutrino).

Taking into account the
recently
revised estimates~\cite{IceCubeNewDirections} of the direction of observation of IceCube neutrinos we only find
(appendix A) 3 GRB-neutrino candidates which appear to be "early", {\it i.e.} neutrinos which could have been sped up by quantum-gravity effects.
And we find (appendix A)
that the probability of finding accidentally at least 3 such early neutrinos in our data sample is a whopping $81 \%$. We conclude that with the revised estimates~\cite{IceCubeNewDirections} of the direction of observation of IceCube neutrinos there is now absolutely no encouragement for the hypothesis of "early" GRB neutrinos.

Next we look at the hypothesis of "late" GRB neutrinos,
where our findings instead are even more intriguing than those previously reported (using erroneous neutrino directions) in Refs.~\cite{Ryan,RyanLensing,NatureAstro,MaNeutriniFirst,HuangLiMaNeutriniNearTeV}.
We find
(appendix B) 7 GRB-neutrino candidates which appear to be "late", {\it i.e.} neutrinos which could have been slowed down by quantum-gravity effects. And we find (appendix B) that the probability of finding accidentally at least 7 such late neutrinos in our data sample is of only $5 \%$.
Evidently the case of late neutrinos deserves further investigation.
Our 7 late GRB-neutrino candidates are shown in Figure 1 and have correlation of $0.56$.

Going back to the background issues that our selection criteria confront us with, it is important to notice that (appendix B) it is actually likely, with probability of $83 \%$, that at least 1 of our 7 late GRB-neutrino candidates be accidental, and the probability  of at least 2 accidental candidates is still a rather high
$39\%$ (while the probability of at least 3 accidental candidates goes down to a more bearable $18\%$). This suggests that, even tentatively assuming the quantum-gravity model is correct, among the data points in Figure 1 it is likely that there are
1 or 2 which play the role of noise.

 We must also stress that in order to produce Figure 1 one must assign a  value of redshift to the GRBs, but only for a minority of GRBs the redshift is known, and in particular only for one of the data points in Figure 1 the redshift was measured.
 As stressed in Refs.~\cite{Ryan,RyanLensing,NatureAstro,MaNeutriniFirst,HuangLiMaNeutriniNearTeV}
 this issue of estimating the redshift for GRBs whose redshift was not measured is rather challenging, also because we can anticipate that the distribution in redshift of GRBs that produce neutrinos will be significantly different from the
 distribution in redshift of
 generic GRBs. When more data will be available the distribution in redshift of GRBs that produce neutrinos could be estimated reliably from the analysis itself: one would use the distribution  of the GRB-neutrino candidates with known redshift to estimate the redshift distribution of
 those whose redshift is unknown.
 With few data points this approach is much less reliable, but nonetheless we follow again
 Refs.~\cite{Ryan,RyanLensing,NatureAstro,MaNeutriniFirst,HuangLiMaNeutriniNearTeV} by estimating the reshift of all long GRBs relevant for Figure 1 on the basis of the single case in which the redshift is known.
 Only one short GRB is relevant for Figure 1 and it is of unknown redshift, so
 (following again
 Refs.~\cite{Ryan,RyanLensing,NatureAstro,MaNeutriniFirst,HuangLiMaNeutriniNearTeV}) we assign to it redshift of 0.6.

 \vspace{-0.2 cm}

\begin{figure}[h!]
	\includegraphics[scale=0.47]{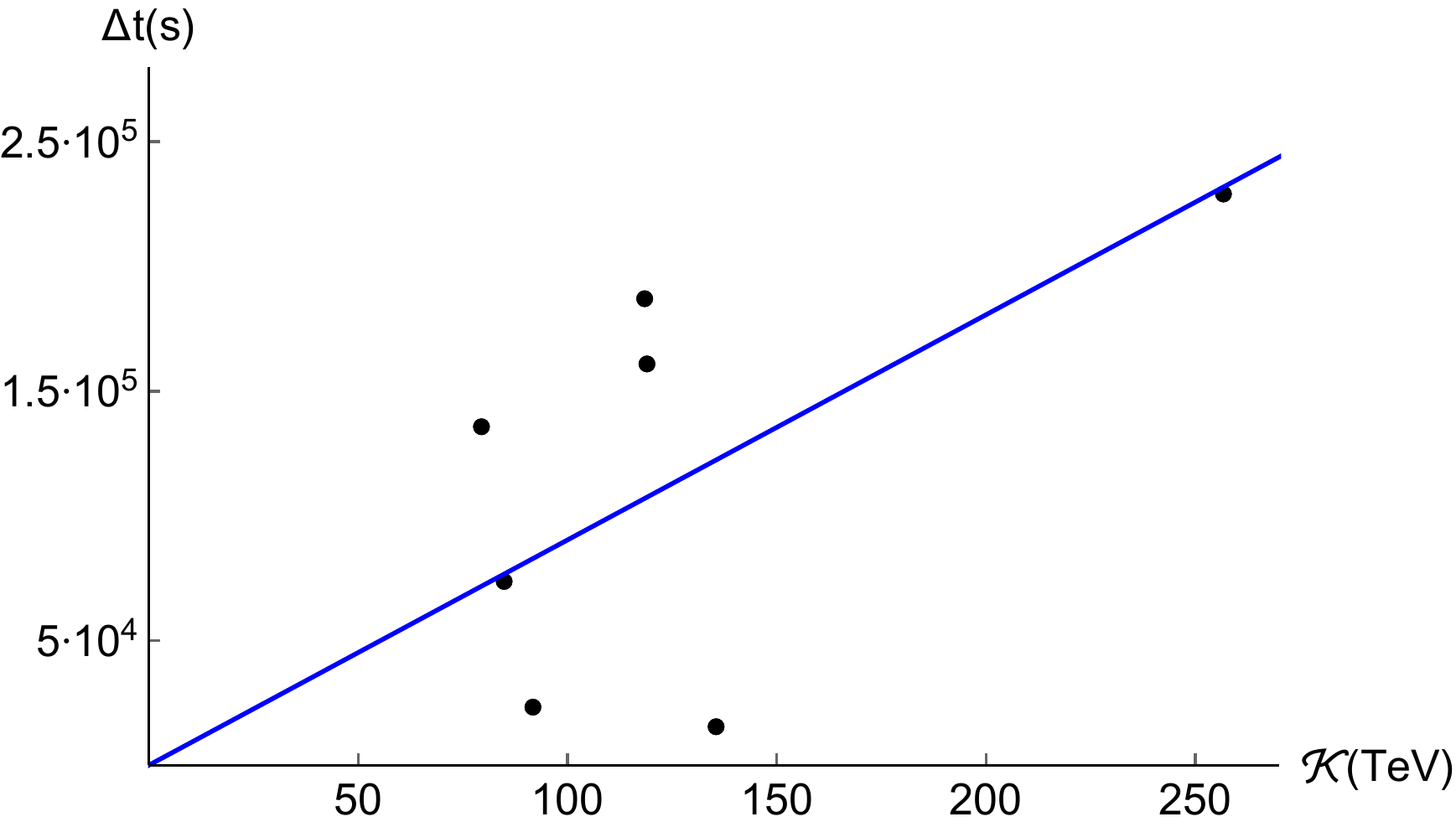}
	\caption{The seven late GRB-neutrino candidates with energy between 60 and 500 TeV described in the main text.
	The blue line was obtained by bestfitting the seven points with the linear relationship between $\Delta t$ and $\mathcal{K}$  given by Eq.(1)}
\end{figure}

\vspace{-0.4 cm}
Also relevant for the understanding of Figure 1 is the fact that for one of our neutrinos
there are three GRBs compatible with it within our temporal and directional window.
The same issue was faced in Refs.~\cite{Ryan,RyanLensing,NatureAstro,MaNeutriniFirst,HuangLiMaNeutriniNearTeV} (with actually more than one neutrino affected by it), and we follow again those previous studies by selecting the GRB which pairs with that neutrino in the way that leads to the highest overall correlation for the analysis.

The line in Figure 1 was obtained by best-fitting the data points, which gives $\eta$=$21.7$ with $\delta \eta$=$4.5$.
Within the model which we are here taking as working assumption the spread of data points around that best-fit line should be interpreted considering that one or two of the points likely are just noise, and for nearly all the points
we relied on a rough estimate of redshift.

Our next task is to estimate
a "false alarm probability", {\it i.e.}
estimate how likely it would be for our data sample to produce accidentally (without any intervening quantum-gravity effects) at least 7 late GRB-neutrino candidates  with correlation of at least  $0.56$.
 Following again what was done in the previous related studies of Refs.~\cite{Ryan,RyanLensing,NatureAstro,MaNeutriniFirst,HuangLiMaNeutriniNearTeV}, we do this by performing $10^5$ randomizations of the times of detection
of the neutrinos relevant for our analysis, keeping their energies and directions fixed,
and for each of these  randomizations we redo the analysis just as if they were real data, including the use of the criterion of selecting highest-correlation cases when multiple GRB partners are found for a neutrino.
We find that this "false alarm probability"
is of only $0.7 \%$, which we feel is our key result for motivating further studies.

Our primary objective has been reached: we established that
the
revised estimates~\cite{IceCubeNewDirections} of the direction of observation of IceCube neutrinos affect  strongly  the quantum-gravity-motivated analysis: there is now no encouragement for the hypothesis of early neutrinos, while, on the contrary, for late neutrinos the preliminary evidence is significantly stronger than found with the previous incorrect estimates of the directions. As we were in the final stages of this study, Ref.~\cite{BlazarSuperluminalLIV} was brought to our attention, and its findings in some sense are consistent with our results: the rather robust evidence~\cite{IceCubeBlazar}
that a
neutrino with energy of at least 183 TeV was observed from the blazar TXS 0506+056 strongly constrains "superluminal" neutrinos (the early neutrinos of our analysis) since such superluminal neutrinos should quickly lose energy via electron-positron pair production~\cite{BlazarSuperluminalLIV}.

In closing, we focus on one additional task: extending the energy range of the analysis above 500 TeV. As stressed in Refs.~\cite{Ryan,RyanLensing,NatureAstro,MaNeutriniFirst,HuangLiMaNeutriniNearTeV}, going below 60 TeV should be useless since most of the additional GRB-neutrino candidates would be background, atmospheric neutrinos. Going above 500 TeV would not pose background problems but it was not done because of the needed size of the time window: the effect grows linearly with energy and therefore, since a 3-day window is needed~\cite{Ryan,RyanLensing,NatureAstro,MaNeutriniFirst,HuangLiMaNeutriniNearTeV} for neutrinos of up to 500 TeV, for a neutrino of, say, 2 PeV, one might have to adopt a 12-day window, in which case the challenge of handling multiple GRB "partners" would grow unmanageably. We here propose a way to include neutrinos with energy greater than 500 TeV that does not require such wide temporal windows: one would still only use the neutrinos with energy between 60 and 500 TeV to estimate a value of the coefficient $\eta$ of Eq.(1) and then look for candidate GRB-neutrinos with energy higher than 500 TeV which are compatible with that estimate of $\eta$.

\noindent
We find (appendix C) two late GRB-neutrino candidates
with energy greater than 500 TeV, which are compatible with $\eta$= $21.7 \pm 9.0 $ (we allowed for a $2 \delta \eta$ interval). Also in this case we encounter the multiple-GRB-partner issue, which again we handle by resorting to the maximum-correlation criterion. The two resulting  GRB-neutrino candidates are shown in red in Figure 2, together with the seven candidates already shown in Figure 1. The overall correlation for the nine points in Figure 2 is a remarkable $0.9997$.

 \vspace{-0.2 cm}

	\begin{figure}[h!]
	\includegraphics[scale=0.47]{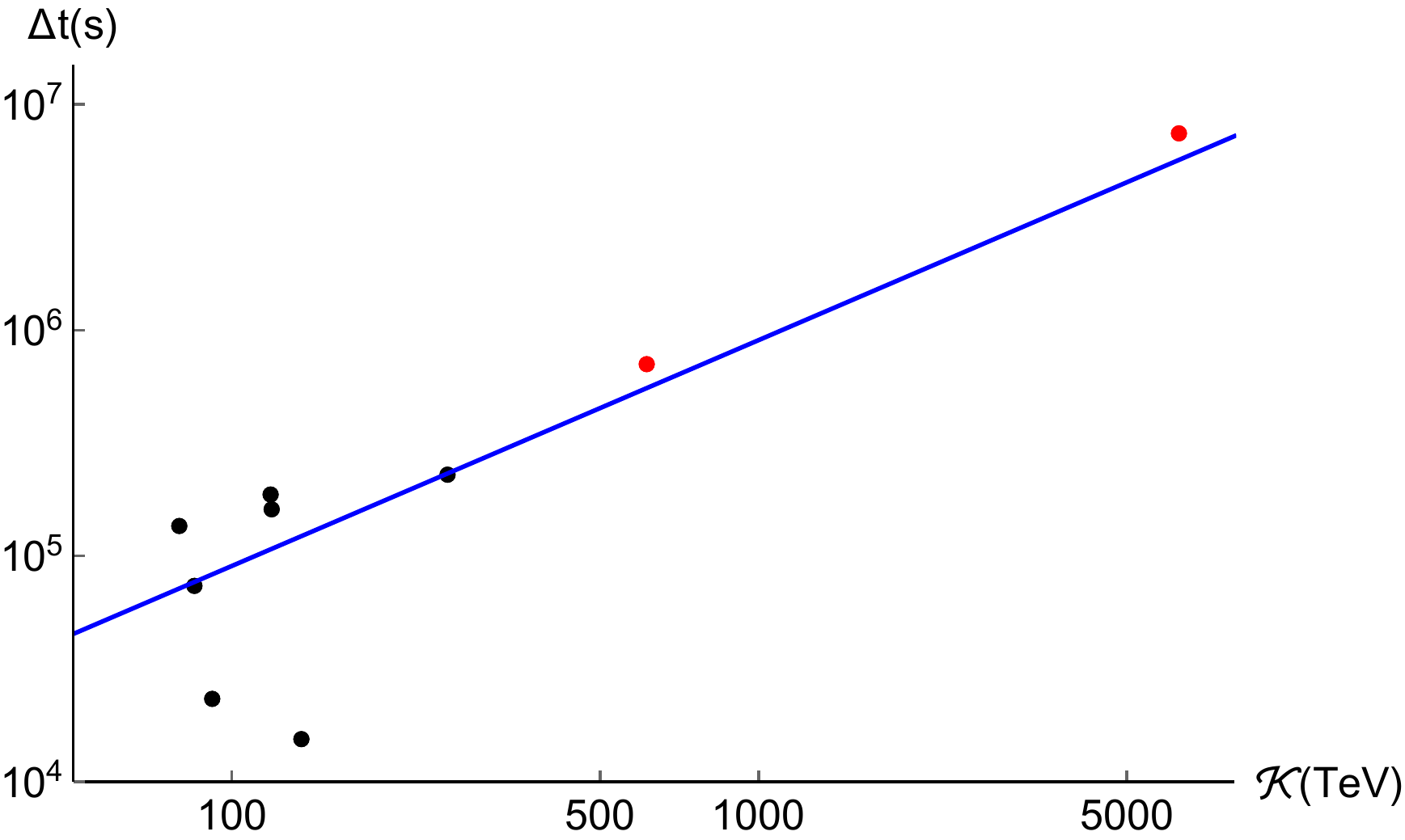}
	\caption{Here the red points are for our two late GRB-neutrino candidates with energy greater than 500 TeV. The black points and the blue line are the same as in Figure 1.}
	\label{im:neutriniPeV}
\end{figure}

 \vspace{-0.4 cm}

This very high value of correlation does not in itself characterize the significance of our findings: also in this case we need a false-alarm probability quantifying how likely it would be for the available neutrinos with energy greater than 500 TeV to accidentally produce late GRB-neutrino candidates leading to this high value of correlation, when combined with our late GRB-neutrino candidates with energy between 60 and 500 TeV.
We performed $10^5$ randomizations of the times of detection
of the available neutrinos with energy greater than 500 TeV, keeping their energies and directions fixed,
and we computed the probability of having at least two late GRB-neutrino candidates with energy greater than 500 TeV  which are compatible with $\eta = 21.7 \pm 9.0$ and produce an overall correlation of at least 0.9997.
Also for our simulated data we handle  the multiple-GRB-partner issue by selecting the case with highest correlation.
We find that the "false alarm probability"
is of only $0.005\%$.

We shall not dwell on the potential significance of this remarkably low false-alarm probability: our investigations started with the goal of studying the impact of Ref.~\cite{IceCubeNewDirections} on analyses focusing on the range 60-500 TeV, and the inclusion of events with energy greater than 500 TeV was an afterthought which, while evidently leading to an intriguing result, we feel should not divert attention from the main part of our analysis. Still, combining our findings in the range 60-500 TeV and our findings for energies greater than 500 TeV,
there appears to be plenty of motivation for monitoring the evolution of this sort of analyses
as more high-energy-neutrino data is accrued.

\section*{Acknowledgements}
G.A.-C.’s work on this project was supported by the
FQXi grant 2018-190483 and by the MIUR, PRIN 2017
grant 20179ZF5KS. G.R.’s work on this project was
supported by the National Science Centre grant
2019/33/B/ST2/00050. This work also falls within the
scopes of the EU COST action CA18108 “Quantum gravity phenomenology in the multi-messenger era”.

\section*{Appendix A: Early neutrinos with energy between 60 TeV and 500 TeV}
\noindent
For the GRBs we use the catalogue that can be found at \url{icecube.wisc.edu/~grbweb_public/Summary_table.html}.
For the neutrinos we use the same data sample of  Refs.~\cite{Ryan,RyanLensing,NatureAstro,MaNeutriniFirst,HuangLiMaNeutriniNearTeV} and we also follow Refs.~\cite{Ryan,RyanLensing,NatureAstro,MaNeutriniFirst,HuangLiMaNeutriniNearTeV} in focusing on "shower events" with energy between 60 TeV and 500 TeV.
Using the selection criteria
of Refs.~\cite{Ryan,RyanLensing,NatureAstro,MaNeutriniFirst,HuangLiMaNeutriniNearTeV} (but now relying on the corrected directional data of the more recent Ref.~\cite{IceCubeNewDirections}) we find three early GRB-neutrino candidates, which are reported in table~\ref{tableearly}.
\begin{table}[htbp]
	\centering
	{\def\arraystretch{0.5}\tabcolsep=3pt
		\begin{tabular}{|c|c|c|r|l|c|c|}
			\hline
		 $E_{\nu}$ \!\!\!\! [TeV]      \!\!\!   &    $  \Delta t$ [s]        & z           & S/L  &   GRB    \\\hline
		  98.5       &  -113050    & -        &    L   &     100605A   \\\hline
			186.6       &  -175141    & -        &    L   &     120224B   \\\hline	
			66.7       &  -234884    & -        &    L   &     140219B   \\\hline	
		\end{tabular}
	}
	\caption{Our three ``early" GRB-neutrino candidates. The first column gives the energy of the neutrinos. The second column gives the difference $\Delta t$ between the time of observation of the neutrino and that of the associated GRB. The third column gives the redshift of the GRB, using "$-$" when the redshift is unkknown (in all these three cases it is unknown). The fourth column specifies whether the GRB is short (S) or long (L). The fifth column specifies the GRB.}
	\label{tableearly}
\end{table}

We estimate the probability that three early GRB-neutrino candidates are found accidentally in our dataset (without any intervening quantum-gravity effect) by generating $10^5$ simulated datasets, each obtained by randomizing the time of observation of the  neutrinos, while keeping their energy and direction fixed, and counting how many times, in these simulated datasets, there are at least 3 neutrinos that find an early GRB-association.
We find that this happens in 81\% of the cases, and we therefore conclude that our three early GRB-neutrino candidates are most likely meaningless, just pure background.

\section*{Appendix B: Late neutrinos with energy between 60 TeV and 500 TeV}
We find seven late GRB-neutrino candidates with energy between 60 TeV and 500 TeV, which we report in table~\ref{tablelate}.
For one of these, with energy of 86.1 TeV, there are three possible associations with a GRB.
Only for the GRB associated with the 61.7 TeV neutrino the redshift is known.

By generating $10^5$ simulated datasets, each obtained by randomizing the time of observation of the  neutrinos, while keeping their energy and direction fixed,
we find that the
 probability
 of obtaining accidentally, without any intervening quantum-gravity effect, at least seven late
 GRB-neutrino candidates with energy between 60 TeV and 500 TeV
 is of only 5\%.

However, even assuming that  some of our seven late
 GRB-neutrino candidates are "signal" ({\it i.e.} their propagation times were affected by quantum gravity)
it is likely that some of them are ''background" ({\it i.e.} they were picked up in the analysis accidentally). In order to see this we perform an analysis which sets aside our seven selected neutrinos:
we randomize the times of observation of the neutrinos which were not selected by our criteria (and therefore should be assumed to be unrelated to any of the GRBs, with or without quantum-gravity effects) and compute how frequently in such randomizations one gets the accidental appearance of late GRB-neutrino candidates. Essentially through these randomizations we estimate which fraction $\zeta$ of an ensemble of neutrinos that surely does not include GRB neutrinos would accidentally be picked up as a GRB-neutrino candidate.
In general, if such an analysis considers $N$ neutrinos and there are $M$ true GRB neutrinos, the number $L$ of GRB-neutrino candidates found clearly will be such that $N \geq L \geq M$, and then one can estimate $M$ (number of GRB-neutrino candidates which actually are GRB neutrinos) through the relationship
$M+\zeta\cdot (N-M)=L$, in which $N$ and $L$ are known, while $\zeta$ is estimated using the randomizations.
Following this procedure we  find a probability of 83\% that at least 1 of our 7 late GRB-neutrino candidates is a background neutrino, a probability of 39\% that at least 2 neutrinos are background,
and
a probability of 18\% that at least 3 neutrinos are background.

Table~\ref{tablelate} also reflects the fact that among the 3 possible GRB partners for the neutrino with energy of 86.1 TeV we picked for Figure 1 the one producing the highest overall correlation.

\begin{table}[htbp]
	\centering
	{\def\arraystretch{0.5}\tabcolsep=3pt
		\begin{tabular}{|c|c|c|r|l|c|c|}
			\hline
		 $E_{\nu}$ \!\!\!\! [TeV]      \!\!\!   &    $  \Delta t$ [s]        & z           & S/L  &   GRB & \! \quad   \\\hline
			98.5       &  15446    & -        &    L   &     100604A  &   *   \\\hline
			86.5       &  160909    & -        &    L   &     110625B & *  \\\hline
			61.7       &  73690    & 1.38        &    L   &     111229A  &  * \\\hline
			86.1       &  200349    & -        &    L   &     120121C  &   \\\hline	
			86.1       &  213239    & -        &    L   &     120121B &    \\\hline	
			86.1       &  187050    & -        &    L   &     120121A &  *   \\\hline	
			186.6       &  229039    & -        &    L   &     120219A  &  *  \\\hline	
			134.2       &  135731    & -        &    S   &     140129C  &  * \\\hline	
			66.7       &  23286    & -        &    L   &     140216A &  *   \\\hline	
		\end{tabular}
	}
	\caption{In this table we report the seven ``late" GRB-neutrino candidates. The first five columns provide the same information as in~\ref{tableearly}. The last column highlights with an asterisk the GRB-neutrino candidates selected by the best-correlation criterion described in the main text.}
	\label{tablelate}
	\end{table}

\section*{Appendix C: Late neutrinos with energy above 500 TeV}
 We select a shower neutrino with energy greater than 500 TeV as late GRB-neutrino candidate if it satisfies the same angular criterion used for the lower energy neutrinos, and if the difference between its time of arrival and that of the GRB is positive and lies within the range $|\Delta t-\eta \cdot \mathcal{K}(E,z) |\leq 2 \delta \eta \cdot \mathcal{K}(E,z)\,.$
We find 2 such late GRB-neutrino candidates with energy greater than 500 TeV, and their properties are given in table~\ref{tablePeV}. For one of these neutrinos we find three possible GRB partners and we handle this again by selecting the case which leads to the highest overall correlation.

	
	\begin{table}[htbp]
		\centering
		{\def\arraystretch{0.5}\tabcolsep=3pt
			\begin{tabular}{|c|c|c|r|l|c|c|}
				\hline
				$E_{\nu}$ \!\!\!\! [TeV]      \!\!\!  &    $  \Delta t$ [s]        & z   & S/L   &   GRB  & \! \quad \\\hline
				1035.5     &  706895    &  -  & S &     110801B  & * \\\hline	
				1035.5     &  907892    &  - & L  &     110730A &  \\\hline	
				1035.5     &  1320217    &  -  & L &     110725A &  \\\hline	
				1800     &  7435884    & 3.93   & L &    120909A & *  \\\hline	
			\end{tabular}
		}
			\caption{Selected GRB PeV-neutrino candidates. The columns provide the same information as in Table~\ref{tablelate}.}
	\label{tablePeV}
	\end{table}

\end{document}